\documentstyle[aps,prl,epsf,preprint]{revtex}
\begin{document}
\widetext
\title{Multicritical crossovers near the dilute Bose gas quantum critical point}
\author{Kedar Damle and Subir Sachdev}
\address{Department of Physics, P.O. Box 208120, Yale University,
New Haven, CT 06520-8120}
\date{\today}
\maketitle
\begin{abstract}
Many zero temperature transitions, involving the deviation in the value of a $U(1)$
conserved charge from a quantized value, are described by the dilute Bose gas quantum
critical point. On such transitions, we study the consequences of
perturbations which break the symmetry down to $Z_N$ in $d$ spatial
dimensions. For the case
$d=1$, $N=2$, we obtain exact, finite temperature, multicritical crossover
functions by a mapping to an integrable lattice model.
\end{abstract}
\pacs{PACS numbers:}

\widetext
The zero temperature ($T$), quantum phase transition in a dilute Bose gas has
recently attracted some interest~\cite{rasolt,fwgf,twod,sss,ferro} because of
its importance in a variety of  different physical situations. 
For bosons with repulsive interactions in a chemical potential, $\mu$, the
quantum critical point is  at $\mu = 0$, where the $T=0$ density of bosons has a
non-analytic dependence on $\mu$. This quantum critical point 
controls the finite $T$ quantum-classical crossovers in the dilute Bose
gas~\cite{rasolt,fwgf,twod,sss}; in addition it is the critical theory for ({\em
i\/}) the Mott-insulator to superfluid transition in a lattice boson system at a
generic $\mu$~\cite{fwgf}, ({\em ii\/}) the onset of
uniform magnetization in a gapped quantum antiferromagnetic in a magnetic
field~\cite{sss}, ({\em iii\/}) the deviation from a saturated polarization in a 
quantum ferromagnet~\cite{ferro}, and possibly other physical
systems~\cite{ferro}.  A common feature of all these systems is that the
Hamiltonian always has at least a global $U(1)$ symmetry, and the transition
involves a deviation in the expectation value of the $U(1)$ conserved charge
from a quantized (possibly zero) value.

In this paper we will study the consequences of perturbations which break the
$U(1)$ symmetry down to $Z_N$ (the cyclic group of $N$ elements), and compute
associated exponents and crossover functions. 
Such an analysis is useful in the experimental spin systems noted above
(especially ({\em ii\/})~\cite{affleck}),
where crystal-induced spin anisotropy destroys the $U(1)$ symmetry. Our most
detailed results will be for $N=2$ and spatial dimension
$d=1$, which is also the most interesting and non-trivial case: we shall
obtain explicit results for multicritical crossovers as a function of $\mu$,
$T$, and the strength of the $U(1)$-breaking perturbation. These are
the first nontrivial, exact results for universal finite $T$ crossovers near a
quantum multicritical point in any system.
 
We begin by reviewing the $U(1)$ symmetric Bose gas theory. The theory is
described by the action
\begin{equation}
{\cal S}_0 = \int d^d x \int_0^{1/T} d\tau \left( \Psi_B^{\dagger}  
\frac{\partial \Psi_B}{\partial
\tau} + \frac{1}{2m} | \nabla_x \Psi_B |^2 -\mu |\Psi_B |^2 + 
\frac{u}{2} |\Psi_B|^4
\right),
\end{equation}
where $\Psi_B$ is a complex scalar field, $m$ is the boson mass and we use
$\hbar=k_B = 1$. A $T=0$ renormalization group analysis shows that the
interaction $u$ is irrelevant for $d>2$~\cite{fwgf}. Near $d=2$, we define the
dimensionless bare coupling $g_0$ by $u=(2m) \kappa^{\varepsilon} g_0 /S_d$
where $\kappa$  is a renormalization momentum scale, $S_d = 2 \pi^{d/2} /((2
\pi)^d
\Gamma(d/2))$ is a phase space factor, and $\varepsilon = 2-d$. The
renormalized theory is defined by a single coupling constant renormalization
of $g_0$ to $g$, and no other renormalizations are necessary. This
renormalization leads to the $\beta$-function
\begin{equation}
{dg}/{d \ln \kappa} = -\varepsilon g + A_d g^2,
\label{betafunc}
\end{equation}
where $A_d = 1/2 + {\cal O}(\varepsilon )$ is a known constant independent of $g$,
with the value the ${\cal O}(\varepsilon )$ terms depending upon
the precise renormalization condition. The result (\ref{betafunc}) is valid to
{\em all\/} orders in $g$. The quantum critical point of ${\cal S}_0$ is at
$\mu=0$, and has dynamic exponent
$z=2$, correlation length exponent
$\nu = 1/2$ for all $d$~\cite{fwgf}. At $T=0$ and $\mu$ small, the boson density $n
= \langle |\Psi_B |^2
\rangle$ obeys~\cite{sss} $n =  \Lambda \theta(\mu) 2 m \mu$ for $d>2$ ($\Lambda$
is a non-universal, cut-off dependent, constant), while for $d<2$ we have
$n = {\cal C}_d \theta(\mu) (2m\mu)^{d/2}$, with ${\cal C}_d$ a
universal number which has non-trivial contributions at each order in
$\varepsilon$~\cite{ferro} due to the fixed-point interaction $g = g^{\ast} =
\varepsilon/A_d$.

We now break the symmetry down to $Z_N$. The most relevant perturbation which
accomplishes this is
\begin{equation}
{\cal S}_N = \int d^d x \int_0^{1/T} d\tau \left( 
\lambda_N \Psi_B^N + c.c.
\right),
\end{equation}
where the coupling $\lambda_N$ can be taken to be real, without loss of
generality. 
For $d>2$, the scaling dimension of $\lambda_N$ is
simply its canonical dimension, $d+2-Nd/2$; 
so $\lambda_2$ is relevant for
all $d>2$ ($\mbox{dim}(\lambda_2)
= 2$), $\lambda_3$ is relevant for $2<d<4$, while $\lambda_{N>3}$
are irrelevant for all $d>2$.  For
$d<2$, we need to consider the renormalization of $\Psi_B^N$ insertions; these are
non-trivial because ${\cal S}_N$ creates or annihilates $N$ bosons, and these
pre-existing bosons can then interact. The two-loop computation of this
renormalization yields the scaling dimensions
\begin{equation}
\mbox{dim}(\lambda_N) = 4 - N -  (N^2 /2 - N + 1) \varepsilon + 
 (N/2)(N-1)(N-2) \ln(4/3) \varepsilon^2+ 
{\cal O}(\varepsilon^3)
\label{epsexp}
\end{equation}
Notice that the ${\cal O}(\varepsilon^2)$ term vanishes for $N=2$. This is also
true for all subsequent terms, as the only renormalization of $\Psi_B^2$ comes
from a single series of ladder diagrams, and we have $\mbox{dim}(\lambda_2) = 2 -
\varepsilon$ to all orders in $\varepsilon$; there is no similar simplification
for $N\geq 3$. Later, we will verify the value of $\mbox{dim}(\lambda_2)$ in $d=1$
by an entirely different method. So
$\lambda_2$ is relevant in
$d=1$. Evaluation of
the series (\ref{epsexp}) at $\varepsilon=1$ predicts that $\lambda_{3 \leq N
\leq 5}$ are irrelevant in $d=1$, while the $\lambda_{N \geq 6}$ are again
relevant. This result for $N\geq 6$ is surely an artifact of the poorer accuracy
of the series at larger $N$, and it is likely that all $\lambda_{N\geq 3}$ are
irrelevant in $d=1$.

The $\lambda_2$ perturbation is expected to drive the quantum 
phase transition in
${\cal S}_0 + {\cal S}_2$ (accessed by the control parameter $\mu$) into the
universality class of the transverse-field Ising model. This latter class has
$z=1$ and upper critical dimension
$d=3$. We will not discuss the details of the crossover between the 
$U(1)$ symmetric, dilute
Bose gas fixed point and the transverse-field Ising fixed point for general $d$
here: we shall confine our attention in the remainder of the paper to the $d=1$
case where {\em both} fixed points are below
their respective upper critical dimensions.

We now present a detailed analysis for $d=1$, $N=2$. 
It has been argued that the $d=1$ critical properties of the action ${\cal S}_0$ 
are identical to those of the $\mu=0$ critical point in a dilute, spinless,
non-interacting, Fermi gas~\cite{sss}. The Bose field, $\Psi_B$, is related to
the Fermi field, $\Psi_F$, by a continuum Jordan-Wigner transformation:
\begin{equation}
\Psi_B (x) = \exp \left( -i \pi \int_{-\infty}^{x}
d x^{\prime} \Psi_F^{\dagger} (x^{\prime} ) \Psi_F (x^{\prime} ) \right)
\Psi_F (x)
\label{jw}
\end{equation}
(we are momentarily interpreting $\Psi_B$, $\Psi_F$ as operators).
We can use this mapping to deduce the mapping of ${\cal S}_2$ into the
fermionic theory. We point split $\Psi_B^2 (x)$ into
$\Psi_B (x) \Psi_B (x + a)$, rewrite in terms of $\Psi_F$ using (\ref{jw}),
and expand in powers of $a$. Retaining only low order terms, we
obtain the following fermionic form for ${\cal S}_0 + {\cal S}_2$ in $d=1$:
\begin{equation}
{\cal S}_F = \int d x \int_0^{1/T} d\tau \left( \Psi_F^{\dagger}  
\frac{\partial \Psi_F}{\partial
\tau} + \frac{1}{2m} | \partial_x \Psi_F |^2 -\mu |\Psi_F |^2
+ \frac{\tilde{\lambda}_2}{2} \left(
\Psi_F^{\dagger} \partial_x \Psi_F^{\dagger} - \Psi_F \partial_x \Psi_F \right)
 \right),
\end{equation}
where $\tilde{\lambda}_2 \propto \lambda_2$, and $\Psi_F$ a
complex Grassmanian field. The action ${\cal S}_F$ is quadratic in fermionic
fields, and correlations of the $\Psi_F$ can be computed. In particular, it can
be shown that all other terms derivable from
${\cal S}_0 + {\cal S}_2$ are irrelevant at the $\mu=0$, $\tilde{\lambda}_2=0$
multicritical point. So we expect the universal, multicritical crossover
functions to emerge from an analysis of ${\cal S}_F$. Note also, from simple
power-counting, that $\mbox{dim}(\tilde{\lambda}_2) = 1$, which agrees with 
our earlier result for $\mbox{dim}(\lambda_2)$, obtained by expanding to all
orders in $\varepsilon$.

\begin{figure}
\epsfxsize=5.8in
\centerline{\epsffile{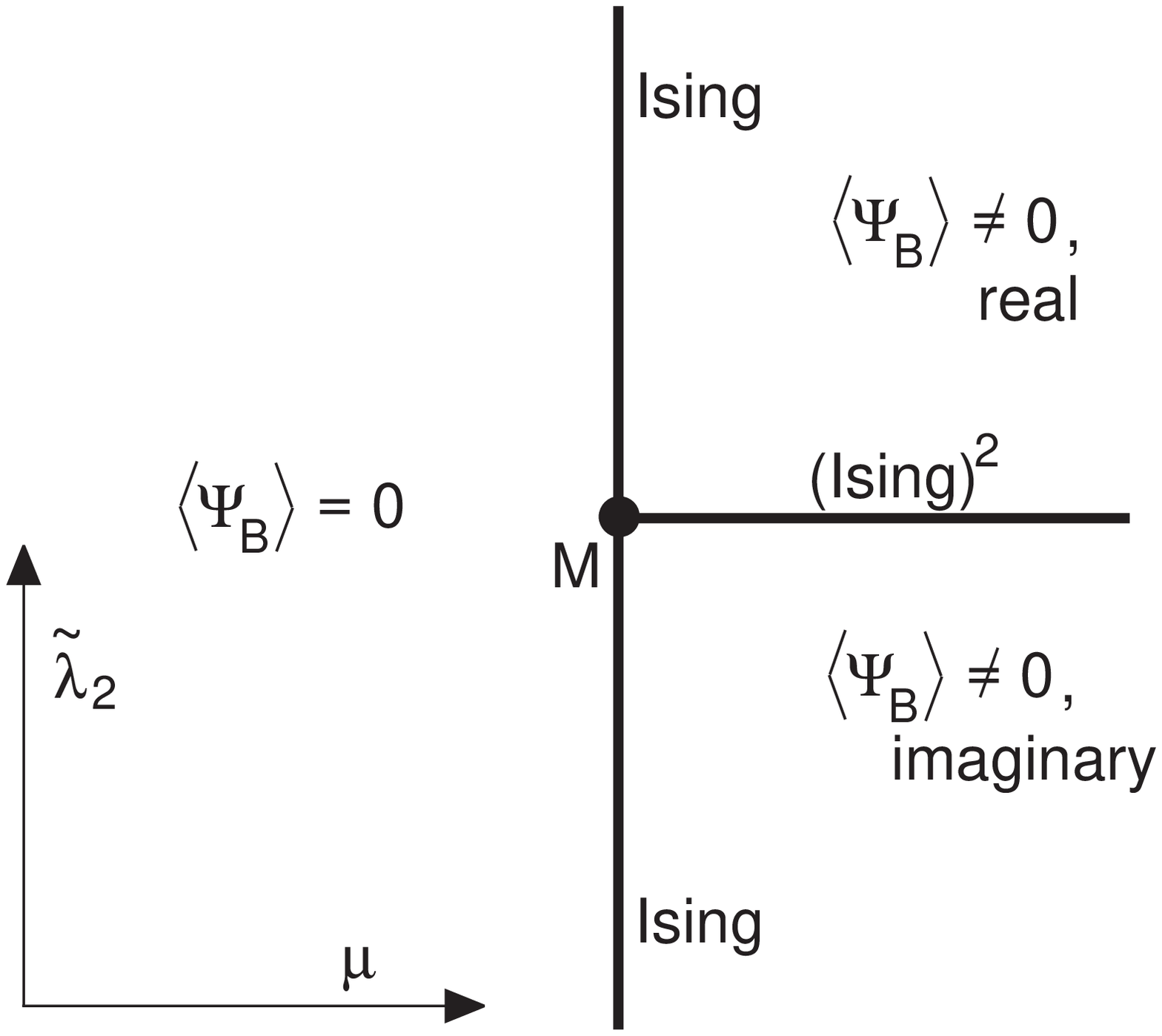}}
\vspace{0.5in}
\caption{Phase diagram of ${\cal S}_F$ (equivalent to ${\cal S}_0 + {\cal
S}_2$ in $d=1$) at $T=0$. 
The lines denote second-order quantum phase transitions with gapless excitations, and
there is gap to all excitations elsewhere. The
multicritical point
$M$ is at
$\mu=0$,
$\tilde{\lambda}_2 = 0$ and has $z=2$. 
The action has $U(1)$ invariance only for $\tilde{\lambda}_2 =0$, and the conserved
$U(1)$ charge, $\langle \Psi_B^{\dagger} \Psi_B \rangle = \langle \Psi_F^{\dagger}
\Psi_F
\rangle$, is quantized at zero for $\tilde{\lambda}_2 = 0$, $\mu < 0$.
On the critical line $\mu > 0$, $\tilde{\lambda}_2 = 0$, this charge varies
continuously and the system is described at sufficiently long scales by the
conformally and $U(1)$ invariant
$(\mbox{Ising})^2$ theory with
$z=1$, central charge $c=1$.
Similarly, the critical lines
$\mu = 0$,
$\tilde{\lambda}_2 >0$ and $\mu=0$, $\tilde{\lambda}_2 < 0$ eventually map
on to the conformally invariant ($z=1$)
Ising field theory with
$c=1/2$. The point $M$ is the only truly scale invariant point, but it is
not conformally (or even Lorentz) invariant.}
\label{f1}
\end{figure}

The $T=0$ phase diagram of ${\cal S}_F$ is shown in Fig.~\ref{f1}.
The $z=2$, dilute Bose gas critical point is $M$, and it is the point of
intersection of three second order phase transition lines; there is a gap to all
excitations everywhere, except at $M$ and along these three lines.
 ({\em i\/})~The
line along
$\tilde{\lambda}_2 = 0$ has
$U(1)$ symmetry, and  describes the  $d=1$ Bose gas with quasi-long range order. At
sufficiently long scales and for $\mu>0$, this line is described by its own critical
theory which has $z=1$, is conformally invariant and has central charge
$c=1$. There is an operator ($|\Psi^{\dagger}_F \partial_x \Psi_F|^2$) which is
marginal along this line, and is responsible for the continuously varying
exponents at $c=1$; however this operator is {\em irrelevant} at the critical
end-point $M$~\cite{sss} and can be ignored while computing the multicritical
crossovers of $M$. Under this condition, the $c=1$ theory can be written
as two copies of the $c=1/2$ Ising field theory. ({\em ii\/}) The lines
$\mu = 0$,
$\tilde{\lambda}_2 >0$ and $\mu=0$, $\tilde{\lambda}_2 < 0$ are also
conformally invariant at long scales, and are then described by a
single $z=1$, $c=1/2$ Ising field theory.
The non-zero expectation values for $\langle \Psi_B \rangle$ (Fig~\ref{f1})
appear only at $T=0$, and we always have $\langle \Psi_B \rangle = 0$ for
$T>0$; this will become clear from our computations below.

We now wish to describe the finite $T$ crossovers in the vicinity of $M$. 
We will study the two-point correlators of $\Psi_B$. It is useful to define
$\Psi_X \equiv \Psi_B + \Psi^{\dagger}_B$ and $\Psi_Y \equiv
-i (\Psi_B - \Psi_B^{\dagger})$. Then, elementary considerations show that, for
$\tilde{\lambda}_2$ real,
$\langle \Psi_X \Psi_Y \rangle = 0$ and $\langle \Psi_X \Psi_X
\rangle_{\tilde{\lambda}_2} = \langle \Psi_Y \Psi_Y \rangle_{-\tilde{\lambda}_2}
$. So it is sufficient to compute $\langle \Psi_X \Psi_X \rangle$ for both
signs of $\tilde{\lambda}_2$.  We will describe the long distance behavior of
its equal time correlation, where we expect
\begin{equation}
\lim_{|x| \rightarrow \infty}
\langle \Psi_X (x, \tau) \Psi_X (0, \tau) \rangle \sim
A e^{-|x|/\xi}~~~~~~\mbox{for $T>0$}.
\label{limpsix}
\end{equation}
The correlation length, $\xi$, and the amplitude, $A$, obey the multicritical
scaling forms
\begin{equation}
\xi^{-1} = \left( 2mT \right)^{1/2} F(x,y)~~~,~~~A=2 \left(2mT \right)^{1/2}
G(x,y)
\label{scale1}
\end{equation}
where $F$ and $G$ are fully universal scaling functions of the dimensionless
variables
\begin{equation}
x = \mu/ T~~~~;~~~~ y = (2m)^{1/2} \tilde{\lambda}_2 /T^{1/2}
\label{scale2}
\end{equation}
The powers of $T$ in (\ref{scale1},\ref{scale2}) follow from the exponents and
scaling dimensions at $M$: $z=2$, $\nu=1/2$, $\mbox{dim}(\mu) = 2$,
$\mbox{dim}(\tilde{\lambda}_2 ) = 1$, and $\mbox{dim} (\Psi_B ) = 1$ (the mass,
$m$, is not to be interpreted as a scaling variable; it converts between the
engineering dimensions of space and time, and is analogous to the
velocity of light in a Lorentz invariant theory). 

\begin{figure}
\epsfxsize=4.8in
\centerline{\epsffile{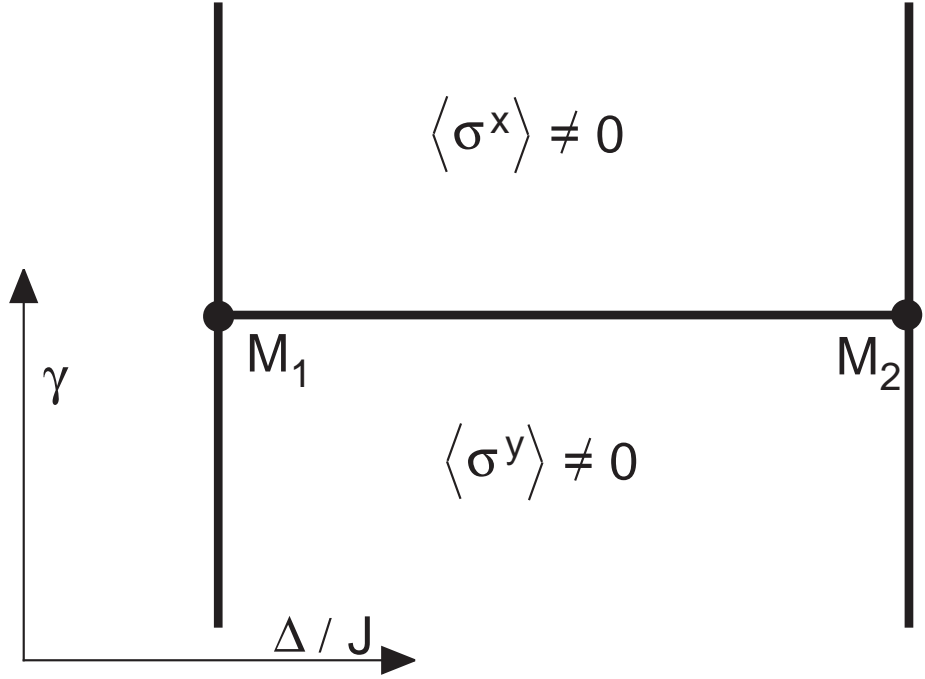}}
\vspace{0.5in}
\caption{Phase diagram of the lattice model $H$ at $T=0$. The expectation
values of $\sigma_x$ and $\sigma_y$ vanish unless otherwise noted. The point
$M_1$ is at $\gamma =0$, $\Delta/J=-2$, while the point $M_2$ is at
$\gamma=0$, $\Delta/J=2$. The vicinities of both $M_1$ and $M_2$ map
independently onto the continuum model of Figure~\protect\ref{f1}.}
\label{f2}
\end{figure}

We will now provide an exact, closed-form, computation of the functions $F$
and $G$. Our strategy is to perform the computation in an integrable lattice model
with a multicritical point in the universality class of $M$. It turns out that
the well-known Lieb-Schultz-Mattis~\cite{lsm} spin chain has two such critical
points. This spin chain is described by the Hamiltonian
\begin{equation}
H = -\sum_{i} \left\{ 
J \left( (1 + \gamma) \sigma^{x}_{i} \sigma^{x}_{i+1} + (1-\gamma)
\sigma^{y}_{i} \sigma^{y}_{i+1} \right) + \Delta \sigma^{z}_{i} \right\}
\end{equation}
where $\sigma^{x,y,z}$ are Pauli matrices on the sites $i$ of an infinite
chain, $J>0$. The Jordan-Wigner transformation maps $H$ into a model of free,
spinless, lattice fermions, and its phase diagram can then be computed
exactly~\cite{lsm,barouch,jimbo}; the result is shown in Fig.~\ref{f2}.
The points $M_1$, $M_2$ are both in the universality class of $M$, and the
continuum limit of the Jordan-Wigner transform of $H$ yields precisely ${\cal
S}_F$; near $M_1$ we find $m = \hbar^2 /(4 J a^2)$, $\tilde{\lambda}_2  = 4
J \gamma a$, $\mu = 2\Delta + 4 J$ ($a$ is the lattice spacing) and the
operator correspondences $\Psi_X \sim \sigma^x$, $\Psi_Y \sim \sigma^y$.
We note that although continuum limits of $H$ have been studied
earlier~\cite{jimbo}, the identification of the universality class of $M_1$,
$M_2$ has not been made.

The required $\Psi_X$ correlators can be obtained from a scaling analysis of
results of Barouch and McCoy~\cite{barouch} on finite $T$ correlators.
Express $\sigma^x$ in terms of the Jordan-Wigner fermions, and evaluate
the resulting fermion correlators; this yields an expression in the form of a
Toeplitz determinant~\cite{lsm,barouch} 
\begin{equation}
\left\langle\sigma^{x}_{i} \sigma^{x}_{i+n} \right\rangle = 
\mbox{det} \left| D_{\ell-m} \right|_{\ell=1\ldots n, m=1\ldots n}
\label{toeplitz}
\end{equation}
where
$D_{\ell} = \int_0^{2 \pi} (d\phi/2 \pi) e^{-i\ell\phi} \tilde{D} (\phi)$
with
$\tilde{D}(\phi) = e^{i\phi} \tanh \left[ | E(\phi)|/k_B T \right]
(E(\phi)/|E(\phi)|)$ and $E(\phi) = 2J \cos \phi + \Delta - i 2 J \gamma \sin
\phi$.
It now remains to take the large $n$ limit of (\ref{toeplitz}); in general,
this is quite difficult, and leads to a computation of considerable
complexity~\cite{barouch}. However, in a limited portion of the phase diagram
($|\Delta/2J| < 1$, $\gamma > 0$ which corresponds to $x>0$, $y>0$ in the scaling
limit associated with $M_1$) the computation is simpler because it is possible to
directly apply  Szego's lemma~\cite{wu}. This limited result is all we shall need as
it is possible to deduce the scaling functions elsewhere by the powerful requirement
that both $F$ and $G$ are analytic for all real, finite $x$ and
$y$~\cite{part1}. This analyticity is a consequence of the absence of
thermodynamic singularities in one dimensional quantum systems at any finite
temperature. The use of analyticity was essential in our being able to express
the final results in a compact form, and it would have been practically
impossible to see the hidden structure in the very lengthy results of
Ref~\cite{barouch} otherwise.

In its region of applicability, Szego's lemma~\cite{wu} tells us
\begin{equation}
\lim_{n \rightarrow \infty} \left\langle\sigma^x_i \sigma^x_{i+n} \right\rangle
\sim e^{n \alpha_0} \exp \left( \sum_{p=1}^{\infty} p \alpha_p 
\alpha_{-p} \right)
\label{szego}
\end{equation}
where
$\ln \tilde{D} (\phi) = \sum_{p=-\infty}^{\infty} \alpha_p e^{ip\phi}$.
We can read off results for $\xi$ and $A$ by comparing (\ref{szego}) with
(\ref{limpsix}). We obtained for the scaling function of the correlation length
\begin{equation}
F(x,y) = \int_0^{\infty} \frac{dq}{\pi}
\ln \coth \frac{ \left[ q^2 y^2 + (x - q^2)^2 \right]^{1/2}}{2}
+ \theta(-y) |y| + \theta(-x) \frac{(y^2 - 4x)^{1/2} - |y|}{2}
\label{resF}
\end{equation}
Only the $x>0$, $y>0$ portion of (\ref{resF}) was obtained from (\ref{szego});
the remainder was deduced by the requirement of analyticity. Indeed, even
though is appears otherwise, the result (\ref{resF}) is in fact a smooth,
differentiable function of
$x,y$ for all real $x,y$ including along the lines $x=0$ and $y=0$.

It is quite interesting to see how the Ising and $(\mbox{Ising})^2$ behavior of
the critical lines in Fig~\ref{f1} emerges from (\ref{resF}). 
First, we observe that along the line $\tilde{\lambda}_2=0$
\begin{equation}
f_B (x) = F(x,0)
\label{limfb}
\end{equation}
where $f_B$ precisely the correlation length crossover function for the $U(1)$
invariant dilute Bose gas in the form presented in~\cite{sss,part1}, where it
was obtained from the results of Ref~\cite{koreppapers}.
The Ising transition realized by crossing the $\mu = 0$ axis for
finite $\tilde{\lambda}_2 >0$ is obtained as follows
\begin{equation}
f_I (x) = \lim_{y\rightarrow \infty} y F(x,y),
\label{limfi}
\end{equation}
where $f_I$ is now the correlation length crossover function of the
transverse-field Ising model obtained in Ref~\cite{part1}. The prefactor of
$y$ on the r.h.s. of (\ref{limfi}) ensures that $f_I$
is multiplied by a power of
$T$ appropriate to the $z=1$ Ising transition.
Crossing $\mu=0$ axis for $\tilde{\lambda}_2 < 0$ requires one to compute
$\langle \Psi_Y \Psi_Y \rangle$ to obtain $f_I$. Finally,
the $(\mbox{Ising})^2$ transition realized by crossing the 
$\tilde{\lambda}_2$ axis for $\mu > 0$ is characterized by the limit
\begin{equation}
f_I (s) = \lim_{x \rightarrow \infty} x^{1/2} F(x, y=s/x^{1/2}).
\label{limfi2}
\end{equation}
Precisely the same function $f_I$ emerges in the very distinct limits in
(\ref{limfi}) and (\ref{limfi2}).

For the amplitude of the correlation function, the computation of $G(x>0, y>0)$
from (\ref{szego})~\cite{barouch} lead to a very lengthy and complicated result.
However, it was found that the analogous result for the Ising
model~\cite{part1} simplified considerably when expressed in terms of derivatives
of the correlation length crossover function. We found the same
remarkable simplification here, and the analytic continuation to all $x,y$ was
then straightfoward; we obtained
\begin{equation}
\log [ G(x,y)] =  - \int_{x}^{\infty} d x^{\prime} \left\{
2 \left( \frac{\partial F(x^{\prime}, y)}{ \partial x^{\prime}} \right)^2
- \frac{1}{x^{\prime}} \left(  y
\frac{\partial F(x^{\prime}, y)}{ \partial
x^{\prime}} + \frac{\partial F(x^{\prime}, y)}{ \partial y} \right)^2 \right\}
- \int_x^{4/y^2} \frac{dx^{\prime}}{4
x^{\prime}} 
\label{resg}
\end{equation}
Both integrands above are singular at $x^{\prime} = 0$, but the singularities
cancel in the sum; it can be verified that 
the expression (\ref{resg}) defines $G$ as function which is smooth for all
real $x,y$, as required. The limiting behavior of $G$ near the critical lines
of Fig~\ref{f1} is similar to that of $F$; we have 
$g_B (x) = G(x,0)$ which is the analog of (\ref{limfb}) (with $g_B$ now the
crossover function of the amplitude of the $U(1)$
invariant dilute Bose gas~\cite{part1,koreppapers}), $\log [g_I (x)] =
\lim_{y\rightarrow
\infty}
\log[ (2/y)^{1/2} G(x,y)]$ as the analog of (\ref{limfi}) for Ising transition line
(with $g_I$ the crossover functions of the amplitude of the Ising
model~\cite{part1}), and $\log [g_I (s)] = (1/2)\lim_{x \rightarrow \infty}
\log[ 2^{1/2} G(x,y=s/x^{1/2})]$ as the analog of (\ref{limfi2}) for the 
$(\mbox{Ising})^2$ transition. Establishing these limits required use of the
recently introduced~\cite{part1} identities obeyed by Glaisher's constant.
 
From the above finite $T$ results for the amplitude, we can also deduce
the value of the spontaneous magnetization: $\lim_{T\rightarrow 0}
A = \langle \Psi_X \rangle^2$ in a regime where $\lim_{T\rightarrow 0}
\xi^{-1} = 0$. This method gives us  a $T=0$ universal function $\Phi$,
\begin{equation}
\langle \Psi_X \rangle =  \theta(\mu) \theta(\tilde{\lambda}_2 ) ( 2m\mu )^{1/4} 
\Phi \left( \frac{(2m)^{1/2} \tilde{\lambda}_2}{\mu^{1/2}} \right)
\end{equation}
with $\Phi (s) = (s/2)^{1/4}$, which describes the crossover of the spontaneous
magnetization between the Ising and $(\mbox{Ising})^2$ lines of Fig~\ref{f1}.

Finally we note that exact, finite temperature, crossover functions of spin
correlators near bulk quantum critical points, below their upper critical dimension,
had previously been computed only for the $d=1$ dilute Bose
gas~\cite{koreppapers,sss,part1} and the
$d=1$ transverse field Ising model~\cite{part1}; our results (\ref{resF}) and
(\ref{resg}) for $F$, $G$ contain these as earlier results as limiting cases, and 
also (universally) interpolate between them. 

This research was supported by NSF Grant DMR-92-24290.


\end{document}